\documentclass[doublecol]{epl2}

\newcommand{\be}{\begin{equation}}\newcommand{\ee}{\end{equation}}
\newcommand{\bea}{\begin{eqnarray}}\newcommand{\eea}{\end{eqnarray}}
\newcommand{\brr}{\begin{array}}\newcommand{\err}{\end{array}}
\newcommand{\bit}{\begin{itemize}}\newcommand{\eit}{\end{itemize}}
\newcommand{\ben}{\begin{enumerate}}\newcommand{\een}{\end{enumerate}}

\newcommand{\ba}{\begin{array}}
\newcommand{\ea}{\end{array}}

\def\lan{\langle}

\def\non{\nonumber}\def\ran{\rangle}

\def\al{\alpha}\def\bt{\beta}

\def\te{\theta}

\def\1{{_{1}}}\def\2{{_{2}}}

\def\noHe0{:\;\!\!\;\!\!:H_e(0):\;\!\!\;\!\!:}
\def\noHm0{:\;\!\!\;\!\!:H_\mu(0):\;\!\!\;\!\!:}

\title{A field-theoretical approach to entanglement in neutrino mixing and oscillations}
\shorttitle{QFT  entanglement in neutrino oscillations} %Insert here a short version of the title if it exceeds 70 characters

\author{Massimo Blasone\inst{1,2} \and Fabio Dell'Anno\inst{2,3} \and Silvio De Siena\inst{2,3,4}
\and Fabrizio Illuminati \inst{2,3,4}}
\shortauthor{M.~Blasone, F.~Dell'Anno, S.~De~Siena and F.~Illuminati}

\institute{
  \inst{1} Dipartimento di Fisica, Universit\`a di Salerno, Via Giovanni Paolo II, 132
84084 Fisciano, Italy
\\
  \inst{2} INFN Sezione di Napoli, Gruppo collegato di Salerno, Italy
\\
\inst{3} Dipartimento di Ingegneria Industriale, Universit\`a   di Salerno, Via Giovanni Paolo II, 132
84084 Fisciano, Italy
\\
\inst{4}
CNISM  Unit\`a di Salerno, I-84084 Fisciano (SA), Italy
}


\abstract{
The phenomena of particle mixing and flavor oscillations in elementary particle physics
can be addressed by the point of view of quantum information theory, and described in terms of multi-mode entanglement of single-particle states. In this paper we show that such a description can be extended to the domain of quantum field theory, where we uncover a fine structure of quantum correlations associated with multi-mode, multi-particle entanglement.
By means of an entanglement measure based on the linear entropies associated with all the possible bipartitions, we analyze the entanglement in the states of flavor neutrinos and anti-neutrinos.
Remarkably, we show that the entanglement is connected with experimentally measurable quantities,
i.e. the variances of the lepton numbers and charges.
}

\begin{document}

\maketitle

\section{Introduction}

In recent years, growing attention has been devoted to the investigation of entanglement in the context of quantum field theory and elementary particle physics~\cite{Bert1,NoiPRD,NoiEPL}. Indeed, understanding the nature of nonlocal quantum correlations in infinite-dimensional systems of fields and particles and their physical consequences poses some subtle conceptual and technical questions.

%For systems composed by identical particles and/or sets of (in general distinguishable) field modes (either discrete or continuous, finite or infinite), it has been clarified that
The proper characterization and quantification of entanglement
%in a mathematically correct form and in a physically sound framework
are achieved unambiguously only by properly taking into account the algebra of observables besides the tensor product structure of the individual state spaces. In the case of quantum fields, this framework must be extended further by considering correlations among distinguishable physical field modes rather than among indistinguishable particles and excitations~\cite{Marinatto,Benatti,Zanardi,Wiseman,Viola,Cirac,VanEnk,Vedral}.
For instance, the single-particle Bell superposition state $|0,1\rangle + |1,0\rangle$ between any two modes of the electromagnetic field is a paramount instance of a maximally (bipartite) entangled quantum state
%(that is with maximal von Neumann entropy of the reduced single-mode density matrices),
%despite the fact that it involves only one excitation of the field (a single photon)
~\cite{VanEnk}.
In this case the entanglement is between two different field modes with occupation numbers ranging between $0$ and $1$.
%Although elementary, this observation has profound physical consequences and is important in order to clarify from the start a confusion that arises often in the community of particle physics.
Considering single-particle or multi-particle (e.g., multi-photon) states of many modes leads to straightforward generalizations that allow to consider the bipartite and multipartite multi-mode field entanglement of single-particle and multi-particle states.
%In the same way, it is possible to make precise sense of Bell nonlocality in the context of single-particle, multi-mode states~\cite{Dunningham2007,Dunningham2009,Heaney2011}.
Among the most spectacular instances of single-particle, multi-mode entanglement, the one associated to particle mixing and flavor oscillations is remarkably outstanding~\cite{NoiPRD,NoiEPL}. Specifically, by considering the physically relevant cases of two- and three-flavor neutrino oscillations, it has been shown that the flavor transition probabilities are simple functions of multi-mode, single-particle entanglement~\cite{NoiEPL}. This observation has concrete operational consequences in that it allows in principle to design experimental schemes for the transfer of the quantum information encoded in neutrino states to spatially delocalized two-flavor charged lepton states~\cite{NoiPRD,NoiEPL}.
%Quantum entanglement is a fundamental resource in quantum information and computation science \cite{NielsenChuang};
%%and topics concerning the study of quantum correlations in paradigmatic quantum systems
%%have been addressed in several branches of condensed matter, atomic physics, and quantum optics .
%a contamination between these sectors and particle physics can turn out both conceptually inspiring and practically useful.
Thus,  entangled states of oscillating neutrinos appears to be  legitimate physical resources for quantum information tasks.
Moreover, the detection of nonlocal quantum correlations in the form of bipartite and multipartite entanglement would provide a decisive test ground for confirming or disproving the quantum nature of fundamental physical phenomena such as particle mixing and flavor oscillations.

By exploiting essential tools of entanglement theory, in the present work we aim at quantifying the content of multi-mode flavor entanglement in systems of oscillating neutrinos. In the course, we will broadly generalize and extend the analysis carried out previously in a simplified, purely quantum-mechanical setting to the framework of Quantum Field Theory (QFT), and we
will develop a fully quantum-field theoretical treatment.
Flavor states of mixed particles have been extensively discussed and a proper QFT framework for their description has already been worked out successfully since some years~\cite{Massimo1,Massimo2,Blasone:2002jv}.
Within such a QFT framework, we  show that the phenomena of flavor oscillations exhibit a richer structure
of quantum correlations with respect to the corresponding quantum mechanical (QM) case previously studied~\cite{NoiEPL}. Indeed, while in QM, flavor neutrino states are single-particle states possessing multi-mode flavor entanglement, in QFT neutrino states are multi-mode flavor-entangled multi-particle states, as the presence of anti-neutrino particle species provides further degrees of freedom
to the oscillating neutrino system.

It is important to remark that entanglement is an observable-induced, relative physical quantity~\cite{Wiseman,Zanardi}, possessing a specific operational meaning according to the reference quantum observables that are singled out in each physical situation, and to the quantum subsystems selected as suitable parties in the partition (bipartite or multipartite) of a physical system.
By assuming the particle-antiparticle species as further quantum modes, we investigate the entanglement content of the  neutrino system in the instance of multi-particle flavor-species entanglement associated with flavor oscillations.
We find that the entanglement quantified via entropic measures is directly related to experimentally measurable quantities,
the variances of the lepton numbers and charges.

The paper is organized as follows:
we first review the quantum information tools exploited in the paper and the main results previously achieved in the simplified QM framework.
We then investigate the entanglement phenomenology of neutrino mixing and flavor oscillations adopting a fully QFT framework,
%we identify the limitations of the simplified QM treatment,
and we discuss the novel, nontrivial structure of flavor entanglement that emerges in the QFT framework.
Finally, we discuss briefly possible outlooks on future research.
%in particular with respect to the experimental detection of flavor entanglement,
%its significance in establishing the quantum nature of particle mixing and flavor oscillations,
%and its possible uses in protocols of quantum information science.

\section{Entanglement and particle mixing: Quantum Mechanics}
\label{QMneutrinoentanglement}

\begin{figure}[t]
\centering
\includegraphics*[width=7.5cm]{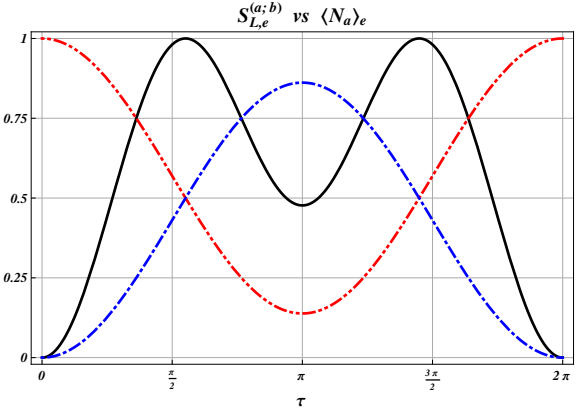}
\caption{(Color online) QM instance: The linear entropies $S_{L,e}^{(a; b)}$
as functions of the scaled time $\tau = (\omega_2 - \omega_1)t$,
with $a,\,b=\nu_e,\,\nu_{\mu}$ (full line), and $a,\,b=\nu_1,\,\nu_2$ (dotted line)
The mixing angle $\theta$ is fixed at the experimental value $\sin^{2}\theta = 0.314$.
The average neutrino flavor-numbers $\langle N_{e} \rangle_e$ (double-dot-dashed line)
and $\langle N_{\mu} \rangle_e$ (dot-dashed line) are also reported.}
\label{FigReview}
\end{figure}

In this section we briefly review the results obtained within the quantum mechanical framework~\cite{NoiPRD,NoiEPL}. In the two flavor case, neutrino mixing is described by the $2 \times 2$
rotation matrix $\mathbf{U}(\theta)$:
\begin{equation}
\mathbf{U}(\theta) = \left( \begin{array}{cc}
  \cos\theta & \sin\theta \\
  -\sin\theta & \cos\theta
\end{array}
\right) \,,
\end{equation}
where $\theta$ is the mixing angle.
The two-flavor neutrino states are defined as
\begin{equation}
|\underline{\nu}^{(f)}\ran \,=\, \mathbf{U}(\theta) \, |\underline{\nu}^{(m)}\ran
\label{fermix2}
\end{equation}
where $|\underline{\nu}^{(f)}\ran \,=\, \left(|\nu_e\ran, |\nu_\mu\ran \right)^{T}$ are the states with definite flavors $e,\mu$
and $|\underline{\nu}^{(m)}\ran \,=\, \left( |\nu_1\ran,|\nu_2\ran \right)^{T}$ those with definite masses $m_1 , m_2$.
Both $|\nu_{\alpha}\rangle$ $(\alpha=e,\mu)$ and $|\nu_{j}\rangle$  $(j=1,2)$ are orthonormal.
By describing the free propagation of the neutrino mass eigenstates
with plane waves of the form $|\nu_j(t)\rangle = e^{-i \omega_{j}t} |\nu_{j}\rangle$,
$\omega_j$ denoting the frequency associated with the mass eigenstate $|\nu_{j}\rangle$,
the time evolution of the flavor states is given by
\begin{equation}
|\underline{\nu}^{(f)}(t)\rangle  = \mathbf{U}(t) |\underline{\nu}^{(f)}\rangle \equiv
\mathbf{U}(\theta)\mathbf{U}_0 (t) \mathbf{U}^{-1}(\theta)|\underline{\nu}^{(f)}\rangle
\label{flavstateevolution}
\end{equation}
where $|\underline{\nu}^{(f)}\rangle$ are the flavor states at $t=0$, and
$\mathbf{U}_{0}(t) = diag (e^{-i \omega_{1}t},e^{-i \omega_{2}t})$.
At time $t$ the average neutrino number of the state $|\nu_{\alpha}(t) \rangle$ in the mode $a$ is:
\begin{equation}
{}\hspace{-.1cm}\langle N_{a}(t)\rangle_{b} \equiv \langle\nu_{b}(t)|N_{a}|\nu_{b}(t)\rangle
=|\mathbf{U}_{b a}(t)|^{2} , \; (a,\,b = e, \mu).
\label{neutrinonumber}
\end{equation}
%
%%%%%%%%%%%%%%%%%%%%%%%%%%%%%%%%%%%%%%%%%%%%%%%%%%%%%%%%%%%%%%%%%%%%%%%%%%%%%%%%%%%%%%%%%%%%%%%%%%%%%%%%%%%%%
By assuming the neutrino occupation number associated with a given mass or flavor (mode) as reference quantum number,
one can establish the following correspondences with two-qubit states:
\begin{eqnarray}
\hspace{-1cm}&&|\nu_{1}\rangle \equiv |1\rangle_{\nu_1} |0\rangle_{\nu_2} \equiv |10\rangle_{12},\;\;
|\nu_{2}\rangle \equiv |0\rangle_{\nu_1} |1\rangle_{\nu_2} \equiv |01\rangle_{12}, \\ [1mm]
\hspace{-1cm}&&|\nu_{e}\rangle   \equiv |1\rangle_{\nu_{e}} |0\rangle_{\nu_{\mu}} \equiv |10\rangle_{e\mu},\;\;
|\nu_{\mu}\rangle \equiv |0\rangle_{\nu_{e}} |1\rangle_{\nu_{\mu}} \equiv |01\rangle_{e\mu},
\end{eqnarray}
where $|j\rangle_{\nu_{\alpha}}$ stands for a $j$-occupation number state of a neutrino in mode $\alpha$.
Entanglement is thus established among mass or flavor modes, in a single-particle setting.
For instance, the free evolution of the electron-neutrino state
$|\nu_e (t)\rangle$ can be written in the alternative forms:
\begin{eqnarray}
|\nu_e (t)\rangle &=& e^{-i\omega_1 t} \cos\theta |10\rangle_{12}  + e^{-i\omega_2 t} \sin\theta |01\rangle_{12} \,,
\label{massBellstate} \\[1mm]
|\nu_e (t)\rangle &=& \mathbf{U}_{e e}(t) |10\rangle_{e\mu}  + \mathbf{U}_{e \mu}(t) |01\rangle_{e\mu} \,,
\label{flavorBellstate}
\end{eqnarray}
where $|\mathbf{U}_{e e}(t)|^2+|\mathbf{U}_{e \mu}(t)|^2=1$.
Thus, the states $|\underline{\nu}^{(f)}(t)\rangle$ can be seen as entangled Bell-like superpositions of
the  mass eigenstates with time-independent coefficients or  of the flavor eigenstates with time-dependent coefficients.
%In particular, the entanglement of Eq.~(\ref{flavorBellstate}) is in principle experimentally accessible,
%throughout a scheme for its transfer from single-neutrino states to two-flavor charged lepton states~\cite{NoiEPL}.
%%%%%%%%%%%%%%%%%%%%%%%%%%%%%%%%%%%%%%%%%%%%%%%%%%%%%%%%%%%%%%%%%%%%%%%%%%%%%%%%%%%%%%%%%%%%%%%%%%%%%%%%%%%%%
The entanglement present in the pure two-qubit states, Eq.~(\ref{massBellstate}) and Eq.~(\ref{flavorBellstate}),
is quantified by the von Neumann entropy of the reduced density matrix, or by any other monotonic function of it, as the linear entropy~\cite{EntRevHorodecki}.
%In view of the subsequent generalization to the multipartite instance,
%a particularly convenient monotonic function of the von Neumann entropy is the linear entropy.
Indeed, such a measure can be suitably exploited to construct measures of multipartite entanglement in multipartite states~\cite{Wootters,Osborne,Adesso2006}, and global measures encompassing both bipartite and multipartite contributions~\cite{BarnumLinden,Wallach,Brennen,Oliveira,WeiGoldbart}.
For example, the global entanglement is associated to the set of linear entropies associated to all possible bi-partitions of the whole system.
We briefly review the definition of the global entanglement, which will be used in this work.
Let $\rho=|\psi\ran\lan \psi|$ be the density operator corresponding to a pure state $|\psi\ran$,
describing the system $S$ partitioned into $N$ parties.
We consider the bipartition of the $N$-partite system $S$ in two subsystems $S_{A_{n}}$,
constituted by $n$ parties $(1\leq n <N)$, and $S_{B_{N-n}}$, constituted by the remaining $N-n$ parties.
Let $\rho_{A_{n}} \equiv Tr_{B_{N-n}}[\rho]$ denote the reduced density matrix of subsystem $S_{A_{n}}$
after tracing over subsystem $S_{B_{N-n}}$.
The linear entropy associated to such a bipartition is defined as
\begin{equation}
S_{L}^{(A_{n};B_{N-n})}(\rho) \,=\, \frac{d}{d-1}(1-Tr_{A_{n}}[\rho_{A_{n}}^{2}]) \,,
\label{linearentropy}
\end{equation}
where the $d$ is the Hilbert-space dimension given by
$d=\min\{\dim S_{A_{n}}\,,\dim S_{B_{N-n}}\}=\min\{2^{n},2^{N-n}\}$.
The corresponding average linear entropy writes:
\begin{equation}
\langle S_{L}^{(n:N-n)}(\rho) \rangle \,=\, \left(%
\begin{array}{c}
  N \\
  n \\
\end{array}%
\right)^{-1}  \sum_{A_{n}} S_{L}^{(A_{n};B_{N-n})}(\rho) \,,
\label{avlinearentr}
\end{equation}
where the sum is intended over all the possible bi-partitions of the
system in two subsystems, respectively with $n$ and $N-n$ elements
$(1\leq n <N)$~\cite{Oliveira}.
%%%%%%%%%%%%%%%%%%%%%%%%%%%%%%%%%%%%%%%%%%%%%%%
The linear entropies Eq.~(\ref{linearentropy})
can be easily computed for the two-qubit Bell state $|\nu_{e}(t)\rangle$,
i.e. Eq.~(\ref{flavorBellstate}), with density matrix
$\rho_e=|\nu_{e}(t)\rangle\langle\nu_{e}(t)|$.
The linear entropies $S_{L,e}^{(\nu_1; \nu_2)}\equiv S_{L}^{(\nu_1; \nu_2)}(\rho_e)$
associated to the reduced state after tracing over one mode (mass) writes:
\begin{equation}
S_{L,e}^{(\nu_1; \nu_2)} =S_{L,e}^{(\nu_2;\nu_1)} =
\sin^{2}2\theta \,,
\label{SLsingletmass}
\end{equation}
which, by fixing $\theta$ at the experimental value $\sin^{2}\theta = 0.314$,
takes the specific value $S_{L,e}^{(\nu_1; \nu_2)} \simeq 0.861$.
The linear entropy $S_{L,e}^{(\nu_e; \nu_{\mu})}\equiv S_{L}^{(\nu_e; \nu_{\mu})}(\rho_e)$
associated to the reduced state after tracing over one mode (flavor) writes:
\begin{eqnarray}
{}\hspace{-.5cm} S_{L,e}^{(\nu_e; \nu_{\mu})} =S_{L,e}^{(\nu_{\mu};\nu_e)}\! &=&\!
4 |\mathbf{U}_{e e}(t)|^{2} \, |\mathbf{U}_{e \mu}(t)|^{2}
\label{SLsinglet}
\end{eqnarray}
(Here and in the following we often omit time dependence for simplicity.)
Thus the linear entropy is strictly related to the average neutrino flavor numbers Eq.~(\ref{neutrinonumber}).
In Fig.~\ref{FigReview} we show the behavior of $S_{L,e}^{(\nu_e; \nu_{\mu})}$
as functions of the scaled, dimensionless time $\tau = (\omega_2 - \omega_1)t$; we also report the neutrino flavor-numbers $\langle N_{e} \rangle_e$
and $\langle N_{\mu} \rangle_e$.
As expected, the entanglement among mass modes, i.e. $S_{L,e}^{(\nu_1; \nu_2)}$, remains constant during the free propagation.
On the contrary, as regards the entanglement among flavor modes, i.e. $S_{L,e}^{(\nu_e; \nu_{\mu})}$, at time $\tau=0$,
the two flavors are not mixed, the entanglement is zero, and the global state of the system is factorized.
For $\tau > 0 $, flavor oscillations occur, and the linear entropy exhibits a typical oscillatory behavior;
the entanglement is maximal at largest mixing, that is $\langle N_{e} \rangle_e=\langle N_{\mu} \rangle_e=0.5$.
We find that the linear entropies $S_{L}^{(a; b)}$ coincide,
apart from a constant factor, to the variances associated with the average neutrino number, i.e.
\begin{eqnarray}
\langle (\Delta N_a )^2 \rangle_e\! &\equiv &\!  \langle N_a^2 \rangle_e - \langle N_a \rangle_e^2   = 4^{-1} \; S_{L,e}^{(a; b)} \,.
\label{VarianceNee}
\end{eqnarray}
Remarkably, the above relation represents a direct experimental connection between entanglement and fundamental physical quantities in the physics of elementary particles \cite{Turchi}.

This observation remains true and is further strengthened when one moves to the QFT setting, as we do in the following Section.
%It is worth noticing that the above results are in complete agreement with the definitions of entanglement in terms of quantum uncertainties
%of physical observables as proposed in Refs.~\cite{Turchi}.

%%%%%%%%%%%%%%%%%%%%%%%%%%%%%%%%%%%%%%%%%%%%%%%%%%%%%%%%%%%%%%%%%%%%%%%%%%%%%%%%%%%%%%%%%%%%%%%%%%%%%%%%%%%%%

\section{Entanglement and particle mixing: Quantum Field Theory}
\label{QFTneutrinoentanglement}

In order to present a generalization of the above analysis to the QFT framework,
we recall the essential features of a specific QFT model of particle mixing
describing the phenomena of neutrino oscillations~\cite{Massimo1,Massimo2}.
The neutrino Dirac fields $\nu_{e}(x)$ and $\nu_{\mu}(x)$ are defined through the mixing relations:
\bea
&&\hspace{-.7cm}\nu_e(x) = \cos\theta \; \nu_1(x) + \sin\theta \; \nu_2(x) \,, \label{nue} \\
&&\hspace{-.7cm}\nu_{\mu}(x) = -\sin\theta \; \nu_1(x) + \cos\theta \; \nu_2(x) \,, \label{numu}
\eea
where, in standard notation, $x$ stands for the four-vector $x \equiv (t,{\bf x})$,
and the free fields $\nu_{1}(x)$ and $\nu_{2}(x)$ with definite masses $m_1$ and $m_2$.
The generator of the mixing transformations is given by:
\begin{eqnarray}
 \hspace{-1cm}&&G_{\bf \te}(t) =
 \exp\left[\theta \int d^{3}{\bf x} \left(\nu_{1}^{\dag}(x) \nu_{2}(x) - \nu_{2}^{\dag}(x) \nu_{1}(x) \right)\right],
 \label{generator12} \\
\hspace{-1cm}&&\nu_{\sigma}^{\alpha}(x) = G^{-1}_{\bf \te}(t)\; \nu_{i}^{\alpha}(x)\; G_{\bf \te}(t),
\end{eqnarray}
where $(\sigma,i)=(e,1),(\mu,2)$, and the superscript $\alpha=1,\ldots,4$ denotes the spinorial component.
At finite volume $V$, $G^{-1}_{\bf \te}(t)$ is a unitary operator,
i.e. $G^{-1}_{\bf \te}(t)=G^{\dag}_{\bf \te}(t)$,
preserves the canonical anti-commutation relations,
and maps the Hilbert space for free fields ${\cal H}_{1,2}$ to
the Hilbert space for mixed fields ${\cal H}_{e,\mu}$, i.e. $G^{-1}_{\bf \te}(t): {\cal H}_{1,2} \mapsto {\cal H}_{e,\mu}$.
In particular, the flavor vacuum is given by
$ |0(t) \rangle_{e,\mu} = G^{-1}_{\bf \te}(t)\; |0 \rangle_{1,2}\; $ at finite $V$.
We denote by $|0 \rangle_{e,\mu}$  the flavor vacuum  at $t=0$.
{It is worth noticing that, in the infinite volume limit,
the flavor and the mass vacua are unitarily inequivalent \cite{Massimo1}, due to the condensate structure
of the flavor vacuum. This also produces a tiny violation of Lorentz invariance
and can be related to dark energy \cite{Capo1,Capo2}. }
The free fields $\nu_{i}(x)$ $(i=1,2)$ are given by the following expansions
\bea\label{freefi}
 \nu _{i}(x)=\frac{1}{\sqrt{V}}{\sum_{{\bf k} , r}}
 \left[ u^{r}_{{\bf k},i}\, \al^{r}_{{\bf k},i}(t) + v^{r}_{-{\bf k},i}\, \bt^{r\dag}_{-{\bf k},i}(t) \right]
 e^{i {\bf k}\cdot{\bf x}},
\eea
where ${\bf k}$ is the momentum vector, $r=1,2$ denotes the helicity,
$\al_{{\bf k},i}^{r}(t)=\al_{{\bf k},i}^{r}\, e^{-i\omega _{k,i}t}$,
$\bt_{{\bf k},i}^{r\dag}(t) = \bt_{{\bf k},i}^{r\dag}\,e^{i\omega_{k,i}t},$
and $ \omega _{k,i}=\sqrt{{\bf k}^{2} + m_{i}^{2}}$.
The operators $\alpha ^{r}_{{\bf k},i}$ and $ \beta^{r }_{{\bf k},i}$ are the annihilation operators
for the vacuum state $|0\rangle_{m}\equiv|0\rangle_{1}\otimes |0\rangle_{2}$,
i.e. $\alpha^{r}_{{\bf k},i}|0\rangle_{m}= \beta ^{r }_{{\bf k},i}|0\rangle_{m}=0$.
For further details see Refs.~\cite{Massimo1,Massimo2}.
By use of $G_{\theta}(t)$, the flavor fields can be expanded as:
\bea\label{flavorfield}
\nu _{\sigma}({\bf x}) =\frac{1}{\sqrt{V}}{\sum_{{\bf k},r} }
\left[ u_{{\bf k},i}^{r}\,  \alpha _{{\bf k},\sigma}^{r}(t) + v_{-{\bf k},i}^{r} \, \beta _{-{\bf k},\sigma}^{r\dagger }(t)\right]
e^{i{\bf k.x}}.
\eea
The flavor annihilation operators are defined as
$\alpha_{{\bf k},\sigma}^r(t)\equiv G_{\theta}^{-1}(t) \alpha_{{\bf k},i}^r G_{\theta}(t)$ and
$\beta_{{\bf k},\sigma}^{r\dagger}(t)\equiv G_{\theta}^{-1}(t) \beta_{{\bf k},i}^{r\dagger} G_{\theta}(t)$.
In the reference frame such that ${\bf k}=(0,0,|{\bf k}|)$,
we have
\bea\non
\hspace{-.7cm}\alpha^{r}_{{\bf k},e}(t)\! &=&\!\cos\theta\;\alpha^{r}_{{\bf k},1}(t)\;
\nonumber \\
&+&\;\sin\theta\;\left( |U_{{\bf k}}|\; \alpha^{r}_{{\bf k},2}(t)\;
+\;\epsilon^{r}\; |V_{{\bf k}}|\; \beta^{r\dag}_{-{\bf k},2}(t)\right),
\label{annihilator}
\eea
etc., with $\epsilon^{r}=(-1)^{r}$ and
\bea
 \hspace{-.7cm} |U_{{\bf k}}| \! & \equiv &\! u^{r\dag}_{{\bf k},i} u^{r}_{{\bf k},j} = v^{r\dag}_{-{\bf k},i} v^{r}_{-{\bf k},j} \,
%\nonumber \\
%\! &=&\!
%\frac{|{\bf k}|^{2} +(\om_{k,1}+m_{1})(\om_{k,2}+m_{2})}{2 \sqrt{\om_{k,1}\om_{k,2}(\om_{k,1}+m_{1})(\om_{k,2}+m_{2})}},
\label{Uk} \\
 \hspace{-.7cm}|V_{{\bf k}}|\! & \equiv & \!  \epsilon^{r}\; u^{r\dag}_{{\bf k},1} v^{r}_{-{\bf k},2}
= -\epsilon^{r}\; u^{r\dag}_{{\bf k},2} v^{r}_{-{\bf k},1}\, ,
%\nonumber \\
%\! &=&\!
%\frac{ (\om_{k,1}+m_{1}) - (\om_{k,2}+m_{2})}{2 \sqrt{\om_{k,1}\om_{k,2}(\om_{k,1}+m_{1})(\om_{k,2}+m_{2})}}\, |{\bf k}| ,
\label{Vk}
\eea
with $i,j = 1,2, \quad i \neq j$, $|U_{{\bf k}}|^{2}+|V_{{\bf k}}|^{2}=1$.
The state of an electron neutrino is defined as $|\nu_{{\bf k},e}^{r} \rangle  \equiv
\alpha_{{\bf k},e}^{r \dag} |0\rangle_{e,\mu}$. In the relativistic limit $|{\bf k}|\gg \sqrt{m_1m_2}\,$, one gets $\left| U_{\mathbf{k}}\right| ^{2}\longrightarrow 1$
and $\left|V_{\mathbf{k}}\right| ^{2}\longrightarrow 0$, and
the corresponding (quantum-mechanical) Pontecorvo state is recovered.
We now omit the superscript $r$ (by fixing $r=2$) and the subscript ${\bf k}$,
thus restricting the analysis to the flavor neutrino state $|\nu_{e} \rangle$ of fixed momentum and helicity.
%Let us consider again the free evolution of the electron-neutrino state (\ref{nuekrt0}):
%\begin{equation}
%|\nu_{e}(t) \rangle = e^{-i H_{0}t}|\nu_{e} \rangle  ,
%\label{nuet}
%\end{equation}
%where $H_{0}$ is the standard QFT free Hamiltonian.
%In the Hilbert space ${\cal H}_{1,2}$, Eq.~(\ref{nuet}) can be written in the form:
In the Hilbert space ${\cal H}_{1,2}$, the neutrino state at time $t$ can be written in the form:
\begin{eqnarray}
|\nu_{e}(t) \rangle \! &=&\! e^{-i\omega_1 t} \big[ \cos\theta \, \alpha_1^{\dag}
+ e^{-i(\omega_2 - \omega_1) t} U \, \alpha_2^{\dag} + \nonumber \\ [1mm]
&& + e^{-i(\omega_2 + \omega_1) t} V \sin\theta \, \alpha_1^{\dag} \alpha_2^{\dag} \beta_1^{\dag} \big] \, |0\rangle_{1,2} .
\label{nuet1}
\end{eqnarray}
In the Hilbert space ${\cal H}_{e,\mu}$, Eq.~(\ref{nuet1}) is rewritten as:
\begin{eqnarray}
|\nu_{e}(t) \rangle \!&=&\! \big[ \mathbf{U}_{ee}(t) \, \alpha_e^{\dag} + \mathbf{U}_{e\mu}(t) \, \alpha_{\mu}^{\dag} +
\mathbf{U}_{e \mu}^{e\bar{e}}(t) \, \alpha_e^{\dag} \alpha_{\mu}^{\dag} \beta_{e}^{\dag} \nonumber \\ [1mm]
&& +\mathbf{U}_{ee}^{\mu\bar{\mu}}(t) \, \alpha_e^{\dag} \alpha_{\mu}^{\dag} \beta_{\mu}^{\dag} \big] \, |0\rangle_{e,\mu} ,
\label{nuet2}
\end{eqnarray}
where the time-dependent coefficients are given by:
\begin{eqnarray}
&&\hspace{-.7cm}\mathbf{U}_{ee}(t) = e^{-i \omega_1 t} \big[ \cos^2\theta +\sin^2\theta
\big(e^{-i(\omega_2-\omega_1)t} |U|^2  \nonumber
\\
&&\hspace{.7cm} +e^{-i(\omega_2+\omega_1)t} |V|^2 \big) \big] , \nonumber
\\[1mm]
&&\hspace{-.7cm}\mathbf{U}_{e\mu}(t) = e^{-i \omega_1 t} U \cos\theta \sin\theta \big(e^{-i(\omega_2-\omega_1)t} -1 \big) , \nonumber
\\[1mm]
&&\hspace{-.7cm}\mathbf{U}_{e\mu}^{e\bar{e}}(t) = e^{-i \omega_1 t} V \cos\theta \sin\theta \big(1 - e^{-i(\omega_2+\omega_1)t} \big) , \nonumber
\\[1mm]
&&\hspace{-.7cm}\mathbf{U}_{ee}^{\mu\bar{\mu}}(t) = e^{-i \omega_1 t} U V \sin^2\theta \big(e^{-i(\omega_2+\omega_1)t} - e^{-i(\omega_2-\omega_1)t} \big), \nonumber
\\[2mm]
&&\hspace{-.7cm} |\mathbf{U}_{ee} (t)|^2 + |\mathbf{U}_{e\mu} (t)|^2 + |\mathbf{U}_{e\mu}^{e\bar{e}} (t)|^2 + |\mathbf{U}_{ee}^{\mu\bar{\mu}} (t)|^2 = 1 .
\end{eqnarray}
In the following, for convenience, we introduce  the real parameters $x = m_2 /m_1$,
$p = |\mathbf{k}| / \sqrt{m_1 m_2}$, and $\tau = (\omega_2 - \omega_1)t$.

%%%%%%%%%%%%%%%%%%%%%%%%%%%%%%%%
Evidently, the $|\nu_e(t)\rangle$ is a multi-particle entangled state both in the mass eigenstates Hilbert space, i.e. Eq.~(\ref{nuet1}),
and in the flavor eigenstates Hilbert space, i.e. Eq.~(\ref{nuet2}).
Analogously with the Pontecorvo states (\ref{fermix2}) and (\ref{flavorBellstate}),
we assume the neutrino occupation number as reference quantum number.
However, with respect to Eqs.~(\ref{massBellstate}) and (\ref{flavorBellstate}), we have still two masses and flavors,
but we have a further degree of freedom that is the neutrino species, i.e. particles and anti-particles.
Therefore, we obtain multipartite entanglement in a four-qubit state.
In the instance of the Hilbert space $\mathcal{H}_{12}$, Eq.~(\ref{nuet1}) can be written as:
\begin{eqnarray}
|\nu_{e}(t) \rangle \!\!&=&\!\! e^{-i\omega_1 t} \cos\theta |1000\rangle_{12\bar{1}\bar{2}} + e^{-i(\omega_2-\omega_1) t} U |0100\rangle_{12\bar{1}\bar{2}} \nonumber \\ [1mm]
 &&+ e^{-i(\omega_2+\omega_1) t} V  |1110\rangle_{12\bar{1}\bar{2}},
\label{nuetImass}
\end{eqnarray}
where $|ijkh\rangle_{12\bar{1}\bar{2}}$ denotes the four-qubit vector
$|i\rangle_{\nu_1} |j\rangle_{\nu_2} |k\rangle_{\bar{\nu}_1} |h\rangle_{\bar{\nu}_2}$ with $i,j,k,h = 0,1$.
On the other hand, in the instance $\mathcal{H}_{e,\mu}$, Eq.~(\ref{nuet2}) can be written as:
\begin{eqnarray}
\hspace{-.7cm} |\nu_{e}(t) \rangle \!&=&\! \mathbf{U}_{ee}(t) |1000\rangle_{e\mu\bar{e}\bar{\mu}} + \mathbf{U}_{e\mu}(t) |0100\rangle_{e\mu\bar{e}\bar{\mu}} + \nonumber \\ [1mm]
&& \mathbf{U}_{e\mu}^{e\bar{e}} (t) |1110\rangle_{e\mu\bar{e}\bar{\mu}}  + \mathbf{U}_{ee}^{\mu\bar{\mu}}(t) |1101\rangle_{e\mu\bar{e}\bar{\mu}} ,
\label{nuetI}
\end{eqnarray}
where $|ijkh\rangle_{e\mu\bar{e}\bar{\mu}}$ denotes the four-qubit vector
$|i\rangle_{\nu_e} |j\rangle_{\nu_{\mu}} |k\rangle_{\bar{\nu}_e} |h\rangle_{\bar{\nu}_{\mu}}$ with $i,j,k,h = 0,1$.
%%%%%%%%%%%%%%%%%%%%%%%%%%%%%%%%%%%%%%%%
Let us analyze the multipartite entanglement possessed by $| \nu_{e} \rangle$, i.e. Eq.~(\ref{nuetI}),
by using the global entanglement defined by Eqs.~(\ref{linearentropy}) and (\ref{avlinearentr}): we compute the linear entropies $S_{L}^{(a;b,c,d)}$
associated with the bipartition of a single-particle subsystem and a three-particle subsystem,
and the corresponding average linear entropy $\langle S_L^{(1:3)} \rangle$.
Next we compute the linear entropies $S_{L}^{(a,b;c,d)}$ associated with the balanced bipartition
of two-particle subsystems, and the corresponding average linear entropy $\langle S_L^{(2:2)} \rangle$.
The linear entropies $S_{L,e}^{(a; b, c, d)}$ associated with Eq.~(\ref{nuetImass}) write:
\begin{eqnarray}
S_{L,e}^{(\nu_1; \nu_2,\bar{\nu}_1,\bar{\nu}_2)} \!&=&\!  4 U^2 \sin^2\theta (1-U^2\sin^2\theta) \,,
\label{SLnuetI1m} \\
S_{L,e}^{(\nu_2; \nu_1,\bar{\nu}_1,\bar{\nu}_2)} \!&=&\!  \sin^2 2\theta \,,
\label{SLnuetI2m} \\
S_{L,e}^{(\bar{\nu}_1; \nu_1,\nu_2,\bar{\nu}_2)} \!&=&\!  4 V^2 \sin^2\theta (1-V^2\sin^2 \theta) \,,
\label{SLnuetI3m} \\
S_{L,e}^{(\bar{\nu}_2; \nu_1,\nu_2,\bar{\nu}_1)}\!&=&\! 0\,,
\label{SLnuetI4m}
\end{eqnarray}
which take specific values for fixed $\theta$, $x$ and $p$.
Of course, in the quantum mechanical limit, Eqs.~(\ref{SLnuetI1m}) and (\ref{SLnuetI2m})
reduces to Eq.~(\ref{SLsingletmass}), while Eqs.~(\ref{SLnuetI3m}) and (\ref{SLnuetI4m}) go to zero.
%By fixing the mixing angle $\theta$ at the experimental value $\sin^{2}\theta = 0.314$, and
%the parameters $x$ and $p$ at $x=10$ and $p=5$, we get:
%\begin{equation}\label{Sstatic}
%\begin{array}{cc}
%  S_{L,e}^{(\nu_1; \nu_2,\bar{\nu}_1,\bar{\nu}_2)} \simeq 0.831,  \quad & S_{L,e}^{(\nu_2; \nu_1,\bar{\nu}_1,\bar{\nu}_2)} \simeq 0.861, \\
%  S_{L,e}^{(\bar{\nu}_1; \nu_1,\nu_2,\bar{\nu}_2)}\simeq 0.075, \quad & S_{L,e}^{(\bar{\nu}_2; \nu_1,\nu_2,\bar{\nu}_1)}=0, \\
%  \langle S_{L,e}^{(1:3)}\rangle \simeq 0.442 . \quad &
%\end{array}
%\end{equation}
The linear entropies $S_{L,e}^{(a; b, c, d)}$
associated with Eq.~(\ref{nuetI}) are given by simple generalizations of Eq.~(\ref{SLsinglet}):
\begin{eqnarray}
S_{L}^{(\nu_{e}; \nu_{\mu},\bar{\nu}_{e},\bar{\nu}_{\mu})} \!&=&\! 4|\mathbf{U}_{e\mu}|^2 (1-|\mathbf{U}_{e\mu}|^2) \,,
\label{SLnuetI1} \\
S_{L}^{(\nu_{\mu}; \nu_{e},\bar{\nu}_{e},\bar{\nu}_{\mu})} \!&=&\! 4|\mathbf{U}_{ee}|^2 (1-|\mathbf{U}_{ee}|^2) \,,
\label{SLnuetI2} \\
S_{L}^{(\bar{\nu}_{e}; \nu_{e},\nu_{\mu},\bar{\nu}_{\mu})} \!&=&\! 4|\mathbf{U}_{e\mu}^{e\bar{e}}|^2 (1-|\mathbf{U}_{e\mu}^{e\bar{e}}|^2)\,,
\label{SLnuetI3} \\
S_{L}^{(\bar{\nu}_{\mu}; \nu_{e},\nu_{\mu},\bar{\nu}_{e})} \!&=&\! 4|\mathbf{U}_{ee}^{\mu\bar{\mu}}|^2 (1-|\mathbf{U}_{ee}^{\mu\bar{\mu}}|^2)\,.
\label{SLnuetI4}
\end{eqnarray}
Of course, in the quantum mechanical limit, Eqs.~(\ref{SLnuetI1}) and (\ref{SLnuetI2}) reduce to the Pontecorvo analogs,
while Eqs.~(\ref{SLnuetI3}) and (\ref{SLnuetI4}) go to zero.

\begin{figure}[t]
\centering
\includegraphics*[width=7.5cm]{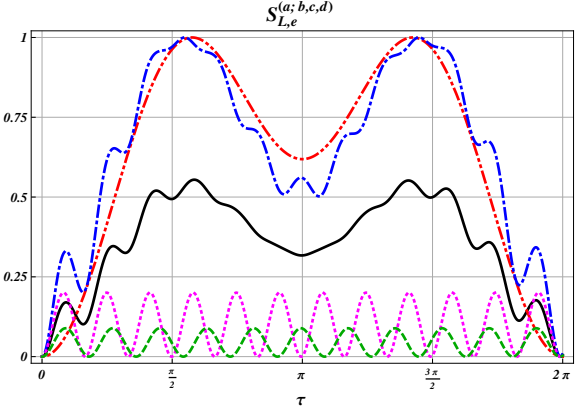}
\caption{(Color online) QFT instance:
The linear entropies $S_{L,e}^{(\nu_e;\nu_{\mu},\bar{\nu}_e,\bar{\nu}_{\mu})}$ (double-dot-dashed line),
$S_{L,e}^{(\nu_{\mu};\nu_e,\bar{\nu}_e,\bar{\nu}_{\mu})}$ (dot-dashed line),
$S_{L,e}^{(\bar{\nu}_e;\nu_e,\nu_{\mu},\bar{\nu}_{\mu})}$ (dotted line),
$S_{L,e}^{(\bar{\nu}_{\mu};\nu_e,\nu_{\mu},\bar{\nu}_e)}$ (dashed line),
and the average linear entropy $\langle S_{L,e}^{(1:3)} \rangle$ (full line)
as functions of the scaled time $\tau = (\omega_2 - \omega_1)t$.
{We take $\sin^{2}\theta = 0.314$, $x=10$ and $p=5$.}
}
\label{FigLEQFTUno}
\end{figure}

\begin{figure}[t]
\centering
\includegraphics*[width=7.5cm]{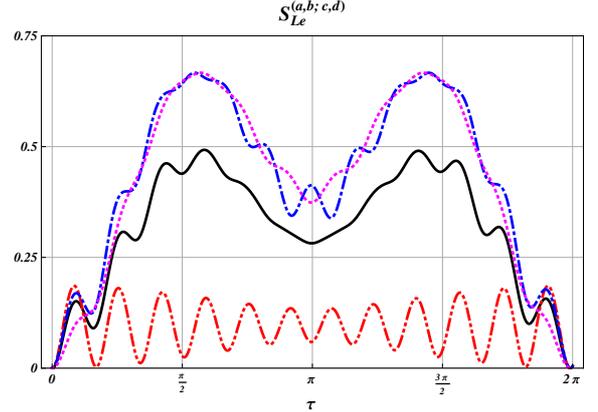}
\caption{(Color online) QFT instance:
The linear entropies $S_{L,e}^{(\nu_e,\nu_{\mu};\bar{\nu}_e,\bar{\nu}_{\mu})}$ (double-dot-dashed line),
$S_{L,e}^{(\nu_e,\bar{\nu}_e;\nu_{\mu},\bar{\nu}_{\mu})}$ (dot-dashed line),
$S_{L,e}^{(\nu_e,\bar{\nu}_{\mu};\nu_{\mu},\bar{\nu}_e)}$ (dotted line),
and the average linear entropy $\langle S_{L,e}^{(2:2)} \rangle$ (full line)
as functions of the scaled time $\tau = (\omega_2 - \omega_1)t$.
The values of parameters are those of {Fig.\ref{FigLEQFTUno}}.
%The mixing angle $\theta$ is fixed at the experimental value $\sin^{2}\theta = 0.314$;
%the parameters $x$ and $p$ are fixed as $x=10$ and $p=5$.
}
\label{FigLEQFTDue}
\end{figure}

In Fig.~\ref{FigLEQFTUno} we plot the quantities (\ref{SLnuetI1})-(\ref{SLnuetI4}) as functions of the scaled time $\tau$ for $x=10$ and $p=5$;
it is worth noticing that, such a choice of the parameters corresponds to the following assumptions:
mass $m_2$ greater than mass $m_1$ of one order of magnitude, and momentum of the same order of magnitude as the masses geometrical mean.
All the linear entropies, except $S_{L,e}^{(\nu_e;\nu_{\mu},\bar{\nu}_e,\bar{\nu}_{\mu})}$,
exhibit rapidly oscillating components, due to the phenomenon of oscillations
involving antineutrinos.
Comparing the QM and QFT instances, i.e. Figs.~\ref{FigReview} and \ref{FigLEQFTUno} respectively,
the linear entropies $S_{L,e}^{(\nu_e; \nu_{\mu})}$ and $S_{L,e}^{(\nu_e; \nu_{\mu},\bar{\nu}_e,\bar{\nu}_{\mu})}$
show a very similar shape and behavior.
%%%%%%%%%%%%%%%%%%%%%%%%%%%%%%%%%%%%%%%%%%%%%%%%%%%%%%%%%%%%%%%%%%%%%%%%%%%%%%%%%%%%%%%%%%%%%%%%%%%%%%%%%%%%%%%%%%%%%%
By straightforward analytical calculation we find that the linear entropies $S_{L}^{(a; b, c, d)}$,
both in the instance of $\mathcal{H}_{12}$ and $\mathcal{H}_{ab}$,
coincide, apart from a constant factor, with the variances associated with the particle number:
\begin{eqnarray}
\langle (\Delta N_{a} )^2 \rangle \!&=&\!\langle N_{a}^2 \rangle - \langle N_{a} \rangle^2 = 4^{-1} \; S_{L}^{(a; b, c, d)} \,.
\label{VarianceNa}
\end{eqnarray}
Therefore, also in the QFT framework, Eq.~(\ref{VarianceNa}) provides a clear operational meaning
for the above entanglement measures throughout a direct connection with measurable physical quantities.
%%%%%%%%%%%%%%%%%%%%%%%%%%%%%%%%%%%%%%%%%%%%%%%%%%%%%%%%%%%%%%%%%%%%%%%%%%%%%%%%%%%%%%%%%%%%%%%%%%%%
Let us now analyze the linear entropies associated with balanced bipartitions.
The linear entropies $S_{L,e}^{(a, b; c, d)}$ associated with Eq.~(\ref{nuetImass}) write:
\begin{eqnarray}
\hspace{-.7cm}S_{L,e}^{(\nu_1, \nu_2;\bar{\nu}_1,\bar{\nu}_2)}\!&=&\! \frac{4}{3} V^2 \sin^2\theta (2-V^2+V^2\cos2\theta) \,,
\label{SLnuetI12m} \\
\hspace{-.7cm}S_{L,e}^{(\nu_1,\bar{\nu}_1; \nu_2,\bar{\nu}_2)} \!&=&\!\frac{2}{3} \sin^2 2\theta \,,
\label{SLnuetI13m} \\
\hspace{-.7cm}S_{L,e}^{(\nu_1,\bar{\nu}_2;\nu_2, \bar{\nu}_1)} \!&=&\! \frac{4}{3} U^2 \sin^2\theta (2-U^2+U^2\cos2\theta) \,
\label{SLnuetI14m} \,.
\end{eqnarray}
%By fixing the mixing angle $\theta$ at the experimental value $\sin^{2}\theta = 0.314$, and
%the parameters $x$ and $p$ at $x=10$ and $p=5$, the above quantities take the values:
%\begin{equation}
%\begin{array}{cc}
% S_{L,e}^{(\nu_1, \nu_2;\bar{\nu}_1,\bar{\nu}_2)} \simeq 0.050, \quad & S_{L,e}^{(\nu_1,\bar{\nu}_1; \nu_2,\bar{\nu}_2)} \simeq 0.574, \\
%  S_{L,e}^{(\nu_1,\bar{\nu}_2;\nu_2, \bar{\nu}_1)}\simeq 0.554, \quad & \langle S_{L,e}^{(2:2)}\rangle \simeq 0.393. \\
%\end{array}
%\end{equation}
The entropies $S_{L,e}^{(a, b; c, d)}$ associated with Eq.~(\ref{nuetI}) write:
\begin{eqnarray}
\hspace{-.7cm}S_{L,e}^{(\nu_{e}, \nu_{\mu};\bar{\nu}_{e},\bar{\nu}_{\mu})} \!&=&\!
\frac{4}{3}\Big[1-(|\mathbf{U}_{ee}|^2+|\mathbf{U}_{e\mu}|^2)^2 \nonumber \\
&&-(|\mathbf{U}_{e\mu}^{e\bar{e}}|^2+|\mathbf{U}_{ee}^{\mu\bar{\mu}}|^2)^2 \Big]  ,
\label{SLnuetI12} \\[1mm]
\hspace{-.7cm}S_{L,e}^{(\nu_{e},\bar{\nu}_{e};\nu_{\mu}, \bar{\nu}_{\mu})} \!&=&\! \frac{4}{3}\Big[1-(|\mathbf{U}_{ee}|^2+|\mathbf{U}_{ee}^{\mu\bar{\mu}}|^2)^2 \nonumber \\
&&-(|\mathbf{U}_{e\mu}|^2+|\mathbf{U}_{e\mu}^{e\bar{e}}|^2)^2\Big]  ,
\label{SLnuetI13} \\ [1mm]
\hspace{-.7cm}S_{L,e}^{(\nu_{e},\bar{\nu}_{\mu};\nu_{\mu}, \bar{\nu}_{e})} \!&=&\! \frac{4}{3}\Big[1-(|\mathbf{U}_{ee}|^2+|\mathbf{U}_{e\mu}^{e\bar{e}}|^2)^2 \nonumber \\
&&-(|\mathbf{U}_{e\mu}|^2+|\mathbf{U}_{ee}^{\mu\bar{\mu}}|^2)^2\Big]  .
\label{SLnuetI14}
\end{eqnarray}
The above quantities are plotted in Fig.~\ref{FigLEQFTDue}.
We observe that, during the time evolution,
most of entanglement lies in the bipartitions $(\nu_e,\bar{\nu}_e;\nu_{\mu},\bar{\nu}_{\mu})$
and $(\nu_e,\bar{\nu}_{\mu};\nu_{\mu},\bar{\nu}_e)$, and, thus, between bipartitions each
containing a particle.
This property also characterizes the linear entropies  $S_{L,e}^{(a; b, c, d)}$, see Fig.~\ref{FigLEQFTUno};
in fact, the unbalanced bipartitions constituted by a single antiparticle-mode versus the remaining modes,
i.e. $(\bar{\nu}_e;\nu_e,\nu_{\mu},\bar{\nu}_{\mu})$ and $(\bar{\nu}_{\mu};\nu_e,\nu_{\mu},\bar{\nu}_{e})$,
possess sensibly less entanglement than the other bipartitions.
Therefore, most entanglement is shared between the two particle-modes.
%%%%%%%%%%%%%%%%%%%%%%%%%%%%%%%%%%%%%%%%%%%%%%%%%%%%%%%%%%%%%%%%%%%%%%%%%%%%%%%%%%%%%%%%%%%%
Finally, it can be analytically demonstrated the existence of a connection between the balanced bipartite linear entropies
and the variances of the neutrino numbers and charges.
Indeed, the following relations hold:
\begin{eqnarray}
 \langle [\Delta (N_{e}+N_{\mu}) ]^2 \rangle \!&=&\! \frac{3}{4} S_{L,e}^{(\nu_e,\nu_{\mu};\bar{\nu}_e,\bar{\nu}_{\mu})} \,,
\label{Var12} \\
 \langle [\Delta (N_{e}-N_{\bar{e}}) ]^2 \rangle \!&\equiv &\! \langle [\Delta Q_e ]^2 \rangle=  \frac{3}{4} S_{L,e}^{(\nu_e,\bar{\nu}_e;\nu_{\mu},\bar{\nu}_{\mu})} \,,
\label{Var13} \\
 \langle [\Delta (N_{e}-N_{\bar{\mu}}) ]^2 \rangle \!&=&\! \frac{3}{4} S_{L,e}^{(\nu_e,\bar{\nu}_{\mu};\nu_{\mu},\bar{\nu}_e)} \,.
\label{Var14}
\end{eqnarray}
In particular, in Eq.~(\ref{Var13}) we have denoted by $Q_e \equiv N_{e}-N_{\bar{e}}$ the charge operator associated with the
electronic flavor. Such an equation clearly refers to the entanglement between flavors independently from the species.
%%%%%%%%%%%%%%%%%%%%%%%%%%%%%%%%%%%%%%%%%%%%%%%%%%%%%%%%%%%%%%%%%%%%%%%%%%%%%%%%%%%%%%%%%%%%

\section{Discussion and Outlook}

In this paper we studied the entanglement associated to neutrino mixing and oscillations in a  field-theoretical framework: the nontrivial nature of QFT flavor states~\cite{Massimo1} gives rise to a complex structure of quantum correlations,
due to the presence of particle-antiparticle pairs. This implies that one deals with multi-mode, multi-particle entangled states.
Characterizing the entanglement by linear entropies, we obtained  exact expressions relating entanglement to experimentally measurable quantities, as the variances of flavor charges. The main observable quantities in particle mixing and flavor oscillations  are thus directly related to entanglement and nonclassicality.

The present analysis has been carried out for the case of two generations.
We plan to extend the study to the case of three flavors, including CP violation, for which the structure of QFT flavor states becomes considerably
more involved~\cite{Blasone:2002jv}. It is worth to remark that the (exact) results obtained in our paper represent a canonical example of evaluation of entanglement for a relativistic system. In this respect, they could be useful for further studies on relativistic properties of entanglement,
e.g. its behavior in connection with Lorentz transformations\footnote{{Lorentz violating effects \cite{Mavro} due to flavor vacuum are
negligible in this context and thus do not affect the above considerations.}} \cite{Friis2009}.
%%%%%%%%

The results of the present analysis, which extend those of ref.\cite{NoiEPL} in the framework of QM to a fully QFT setting, may be exploited to investigate quantum information protocols based on oscillating neutrinos, to be realized in analogy to the use of mode entanglement of the electromagnetic field, in particular those protocols based on single-photon entanglement. Concerning experimental tests and implementations, entanglement can be in principle experimentally accessed via a transfer scheme from single-neutrino states to two-flavor charged lepton states~\cite{NoiEPL}.
%Alternatively, one might consider a generalization to single-neutrino systems of experiments that determine single-particle entanglement indirectly by evaluating the output-input fidelity in experiments on deterministic teleportation using single-photon entanglement as the nonclassical shared resource~\cite{Bjork2011}.
Finally, one could devise extensions to single-neutrino systems of the experimental schemes  based on single-neutron interferometry
that were able to detect {contextuality} in single-nucleon systems~\cite{Rauch2003}.
{In the instance of three generations, it would be interesting to investigate the multipartite entanglement and
to devise an experimental protocol for its detection similar to that of Ref.~\cite{Erdosi}.}

\acknowledgments
The authors acknowledge support from the EU STREP Project iQIT, Grant Agreement No. 270843.

\end{document}